\documentclass[times,twocolumn,final]{elsarticle}

\usepackage{prletters}
\usepackage{framed,multirow}

\usepackage{amssymb}
\usepackage{latexsym}

\usepackage{url}
\usepackage{xcolor}
\definecolor{newcolor}{rgb}{.8,.349,.1}

\journal{Pattern Recognition Letters}

\begin{document}

\thispagestyle{empty}

\ifpreprint
  \setcounter{page}{1}
\else
  \setcounter{page}{1}
\fi

\begin{frontmatter}

\title{MISS GAN: A Multi-IlluStrator Style Generative Adversarial Network for image to illustration translation}

\author[1]{Noa Barzilay\corref{cor1}} 
\cortext[cor1]{Corresponding author: noabarzilay11@gmail.com}
\author[1]{Tal Berkovitz Shalev}
\author[1]{Raja Giryes}

\address[1]{School of Electrical Engineering, Tel Aviv University, Tel Aviv 69978, Israel}

\received{1 May 2013}
\finalform{10 May 2013}
\accepted{13 May 2013}
\availableonline{15 May 2013}
\communicated{S. Sarkar}

\begin{abstract}
Unsupervised style transfer that supports diverse input styles using only one trained generator is a challenging and interesting task in computer vision. This paper proposes a Multi-IlluStrator Style Generative Adversarial Network (MISS GAN) that is a multi-style framework for unsupervised image-to-illustration translation, which can generate styled yet content preserving images. The illustrations dataset is a challenging one since it is comprised of illustrations of seven different illustrators, hence contains diverse styles. Existing methods require to train several generators (as the number of illustrators) to handle the different illustrators' styles, which limits their practical usage, or require to train an image specific network, which ignores the style information provided in other images of the illustrator. MISS GAN is both input image specific and uses the information of other images using only one trained model.\newline
Keywords: Generative Adversarial Networks, Image to Image Translation, Multi-style Transfer.
\end{abstract}

\begin{keyword}
\MSC 41A05\sep 41A10\sep 65D05\sep 65D17
\KWD Generative Adversarial Networks\sep Image to Image Translation\sep Illustration\sep Multi Style Transfer 

\end{keyword}
\end{frontmatter}

\setcitestyle{square}
\section{Introduction}
\label{sec1}
Many works have demonstrated the power of neural networks in creating new images with artistic styles in the image-to-image style transfer task following Gatys et al. \cite{gatys2016neural}. In this task, given a source content image from one domain and a style reference image from another domain, the generated image should resemble the style of the reference image while maintaining the semantic content of the source image. Style transfer can be used in various computer vision applications such as artistic image generation \cite{huang2017arbitrary, xu2018learning}, natural images to painting transfer \cite{zhu2017unpaired}, data augmentation \cite{jackson2019style}, transferring real-world scenes into cartoon style \cite{chen2018cartoongan}, controlling features of a face image using another face \cite{zeno2020ctrlfacenet}, transforming certain features of the scene \cite{huang2018multimodal}, etc.

\begin{figure*}[!t]
\centering
\includegraphics[width = 1\linewidth]{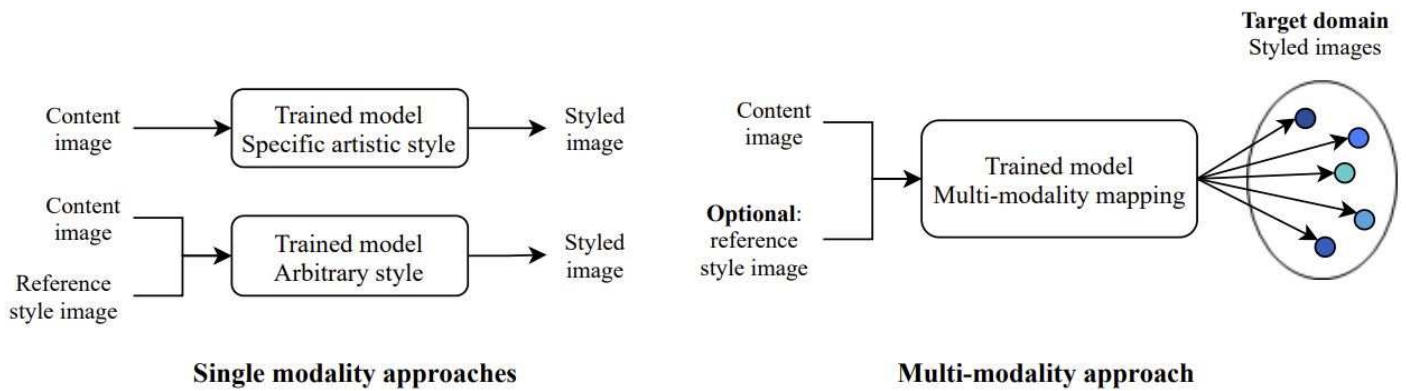}
\caption{Single modality mapping approaches compared to multi-modality mapping approaches. On the left: two approaches for single modality mapping. The first is style transfer for a specific artistic style and the second is generating an arbitrary artistic style given a pair of content and style images. On the right: the multi-modality mapping approach that receives a content image and outputs multiple images of diverse artistic styles.}
\label{fig:ModalityVsMultiModal}
\end{figure*}

In various cases, mapping one domain to another is a \emph{multi-modal} task that entails learning an arbitrarily great number of diverse styles. 
In this work, the term \emph{modality} refers to a specific style in one domain, i.e., a specific artistic style in the illustration domain. Thus, the concept of \emph{multi-modality} style transfer mapping refers to the task of learning to transfer from one domain to another with multiple diverse styles (see Fig.~\ref{fig:ModalityVsMultiModal}).
For example, in mapping a given edge map image to a realistic shoe image, the generated shoe image can have a variety of different fashion design styles \cite{huang2018multimodal}, which may be related to its designer that usually uses similar shoe materials, colors, patterns, etc. Thus, having the designer information when transferring the style of a given image should improve results. 

In this study, transferring an image from a source domain to the target domain is done under the assumption that the mapping is multi-modal in nature. 
This task is challenging since most style-transfer networks are uni-modal and thus will mix information between different modalities or use only a single image as input and thus miss the information available from other images (especially the ones of the same modality).

Designing models that learn multi-modal style mapping may be complicated due to the diversity of styles that can be found in a given dataset. Several works have developed GAN-based multi-modal image-to-image translation \cite{huang2018multimodal, choi2020stargan, sendik2020unsupervised}. 
This work develops a model that maps natural images to illustrations in children's books. It uses only one generator that is able to learn the various illustrators' styles available and is input specific. 

As we show hereafter, the existing multi-modal solution does not give satisfying results for this problem and existing approaches for children's books illustrations  require using a different model per illustrator \cite{ hicsonmez2020ganilla}. 


{\bf Contribution.} To address the style diversity, we propose a novel framework which can learn multi-modality mapping for children's books illustrations. Instead of training a generator for each illustrator, we introduce a new generator based on the GANILLA generator \cite{hicsonmez2020ganilla}, which is combined with an existing multi-modal state-of-the-art architecture called StarGAN v2 \cite{choi2020stargan}. We enhance the GANILLA generator by adding residual blocks to its decoder and show that its incorporation in StarGAN v2 improves the results. Moreover, to achieve a further improvement, we propose adding a feature content objective function to the StarGAN v2 framework. Each of these contributions is validated with a corresponding ablation study. We further demonstrate the importance of replacing the generator in the StarGAN v2 framework architecture for the illustrations dataset. Additionally, we utilize a dataset that consists of 7 of the illustrators that were presented in \cite{hicsonmez2020ganilla} with an increased amount of illustrations (see Table \ref{table:IlluTab}).

\section{Related Work}
GANs received wide attention in recent years since Goodfellow et al. \cite{goodfellow2014generative} pioneering work. They achieved impressive results in various applications such as text to photo-realistic image synthesis \cite{ak2020semantically, qi2021mrp}, super-resolution \cite{ledig2017photo}, face synthesis \cite{huang2017beyond}, colorization \cite{yoo2019coloring}, etc. Karras et al. \cite{karras2019style} introduced StyleGAN, which integrates the Adaptive Instance Normalization (AdaIN) style control mechanism \cite{huang2017arbitrary} that provides some control over attributes at different scales. 

In recent years, several approaches for image-to-image style transfer were developed. Part of them concentrated on learning input-based style transfer, i.e., learning the style of a specific input image \cite{gatys2016image, johnson2016perceptual}. Gatys et al. \cite{gatys2016image} developed a neural framework for image-to-image style transfer, where deep features were used to represent the content, and a Gram matrix was used to represent the style. Johnson et al. \cite{johnson2016perceptual} introduced the perceptual loss, which calculates the distance in the features space of a pre-trained VGG network \cite{simonyan2014very} between the output and the reference image. This loss was proven to be efficient for the image-to-image style transfer task. 

Another strategy focused on learning arbitrary artistic style given a set of content and style images \cite{huang2017arbitrary, xu2018learning, zhang2017multi, sanakoyeu2018style, li2017universal}. Yet, those works generate an output that is solely dependent on the input, i.e., they do not support the generation of multiple diverse styles for a given input. Huang et al. \cite{huang2017arbitrary} proposed a network that uses AdaIN layers to control the style of the content image. In \cite{zhang2017multi}, a CoMatch layer was proposed to match the second order
feature statistics with the target styles. Xu et al. \cite{xu2018learning} introduced a feed-forward network with encoder-decoder architecture as the generator for arbitrary style transfer. In \cite{li2017universal}, whitening and coloring transforms were proposed in the deep feature space to directly match content feature statistics to those of the style reference image. Sanakoyeu et al. \cite{sanakoyeu2018style} presented style aware content loss for real-time HD style transfer, which significantly improved stylization by capturing how style affects content. Another approach tried to tackle the style diversity by learning a multi-modality mapping from the source domain to the target domain \cite{huang2018multimodal, choi2020stargan, zhu2017toward, lee2018diverse}. Zhu et al. \cite{zhu2017toward} explored several different training objectives and network architectures for achieving multiple possible outputs in the image-to-image translation task. In \cite{lee2018diverse}, an approach based on disentangled representation was presented for producing diverse outputs. Huang et al. \cite{huang2018multimodal} developed an architecture called MUNIT that assumes a partially shared latent space. Their architecture contains an auto-encoder that consists of a content encoder, style encoder, and decoder that uses a multilayer perceptron (MLP) to produce a set of AdaIN parameters from the output of the style encoder. The difference between these multi-modality mapping approaches and single modality methods is visualized in Fig. \ref{fig:ModalityVsMultiModal}.

\begin{figure*}[!t]
\centering
\includegraphics[width = 1\linewidth]{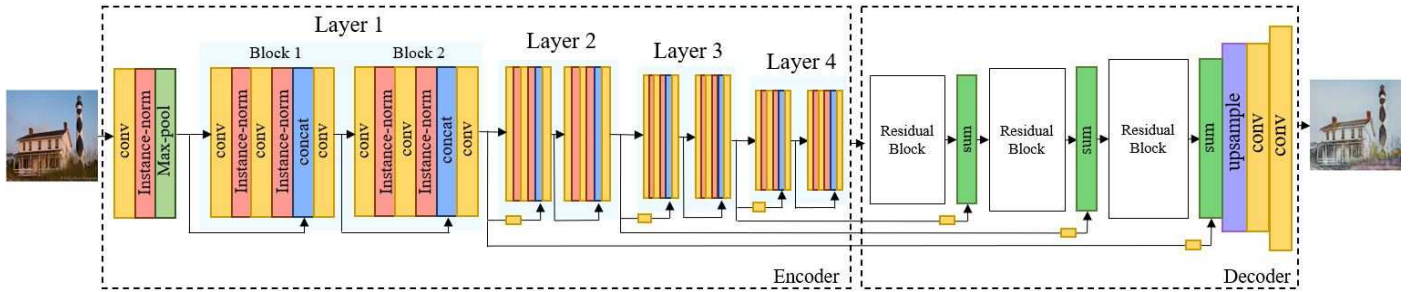}
\caption{MISS GAN Generator. The encoder of the generator is similar to the GANILLA encoder architecture \cite{hicsonmez2020ganilla}. The decoder of the generator contains three residual blocks. The architecture of each residual block is shown in Fig. \ref{fig:ResDecoderBlock}. }
\label{fig:MISSGANGen}
\end{figure*}

\begin{figure}[!t]
\centering
\includegraphics[scale=.78]{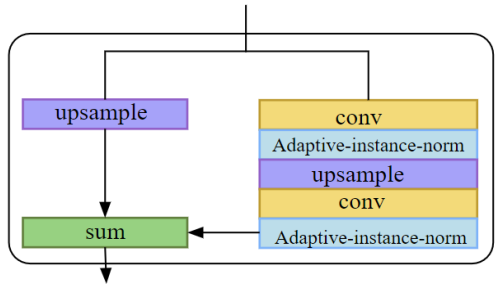}
\caption{Overview of the architecture of the residual blocks that are used in the MISS GAN generator's decoder}
\label{fig:ResDecoderBlock}
\end{figure}

Recently, Choi et al. \cite{choi2020stargan} introduced the StarGAN v2 framework, where one of its main goals is producing images with diverse styles of a specific domain, hence, learning a multi-modal distribution that represents the target domain. Their network achieved unprecedented results and is now considered a state-of-the-art architecture for generating diverse images across various domains. It consists of four main components: a discriminator, a generator, and additional two components that create the style code: The mapping network that generates it from a randomized latent code, and the style encoder that extracts it from a reference image. In this study, we adopt the StarGAN v2 multi-modal framework \cite{choi2020stargan} and combine it with an improved version of the GANILLA generator in order to create a generator that is capable of learning diverse illustrators' styles.

\section{The MISS GAN model}

We turn to introduce now our proposed Multi-IlluStrator Style Generative Adversarial Network (MISS GAN) for image to illustration translation. It combines the StarGAN v2 framework \cite{choi2020stargan}, which allows learning multi-style transfer and performing an unsupervised image-to-illustration translation, with a novel generator which is based on the GANILLA generator \cite{hicsonmez2020ganilla}. This leads to achieving improved and more diverse illustrations that are both appealing and content preserving.

The StarGAN v2 \cite{choi2020stargan} framework consists of four modules: generator, mapping network, style encoder, and discriminator. The generator receives the AdaIN information either from the style encoder that receives a reference image or from the mapping network that receives a latent code. 
We replace this generator with a modified version of the GANILLA generator. The resulting MISS GAN architecture is presented in Fig. \ref{fig:MISSGANGen} and the residual block is detailed in Fig.~\ref{fig:ResDecoderBlock}). Note that the encoder of the generator is similar to the GANILLA generator, which starts with a 7x7 convolution layer, followed by an instance normalization layer \cite{ulyanov2016instance}, ReLU, and max pooling layers. Then, the generator continues with four layers, where each layer consists of two residual blocks. As opposed to  the architecture presented in \cite{hicsonmez2020ganilla}, our decoder contains three residual blocks (Fig. \ref{fig:ResDecoderBlock}). Each residual block starts with a convolution layer, followed by an AdaIN layer and ReLU activation. These are followed by a simple upsample layer, convolution layer, AdaIN layer, and ReLU activation. The use of AdaIN in our model makes it input dependent, which is a very desirable property in style transfer.

\subsection{Training objectives}
Let $X$ and $Y$ be two different domains, where $x \in X$ is an image from the first domain, and $y \in Y$ is an image from the second domain. With these notations, we turn to describe the objective functions tested.


As a baseline we use the same loss functions that are employed in the StarGAN v2 framework \cite{choi2020stargan}. The first objective is the adversarial objective ($\mathcal{L}_{adv}$) which is computed by using two discriminator predictions corresponding to the input domain and target domain over the input image and the generated image respectively. The generated image in the adversarial objective was created by using the style code that was produced in the mapping network. The adversarial objective ensures that the mapping network learns to offer the style code that is likely to be taken from the target domain. Moreover, it guarantees that the generator will learn to utilize this style code and will produce an image that is indistinguishable from other images of the target domain. The second objective is the style reconstruction objective ($\mathcal{L}_{sty}$) which enforces the generator to utilize the style code that was created in the mapping network component while receiving a generated latent code. This objective is calculated using the style encoder output over the generated image. The third objective is the style diversification objective ($\mathcal{L}_{ds}$) which is calculated by two different generated images that were created from two different style codes. The style codes for the style diversification objective were obtained by generating two latent codes and passing them forward to the mapping network component. This objective further enables the generator to produce diverse images. The fourth objective is for preserving the source characteristics of the input image ($\mathcal{L}_{cyc}$ which is essentially the cycle consistency loss \cite{zhu2017unpaired}). 

In addition to the above four losses, we explore the impact of adding two other loss functions to further boost performance.

\begin{figure*}[!t]
\centering
\includegraphics[width = \linewidth]{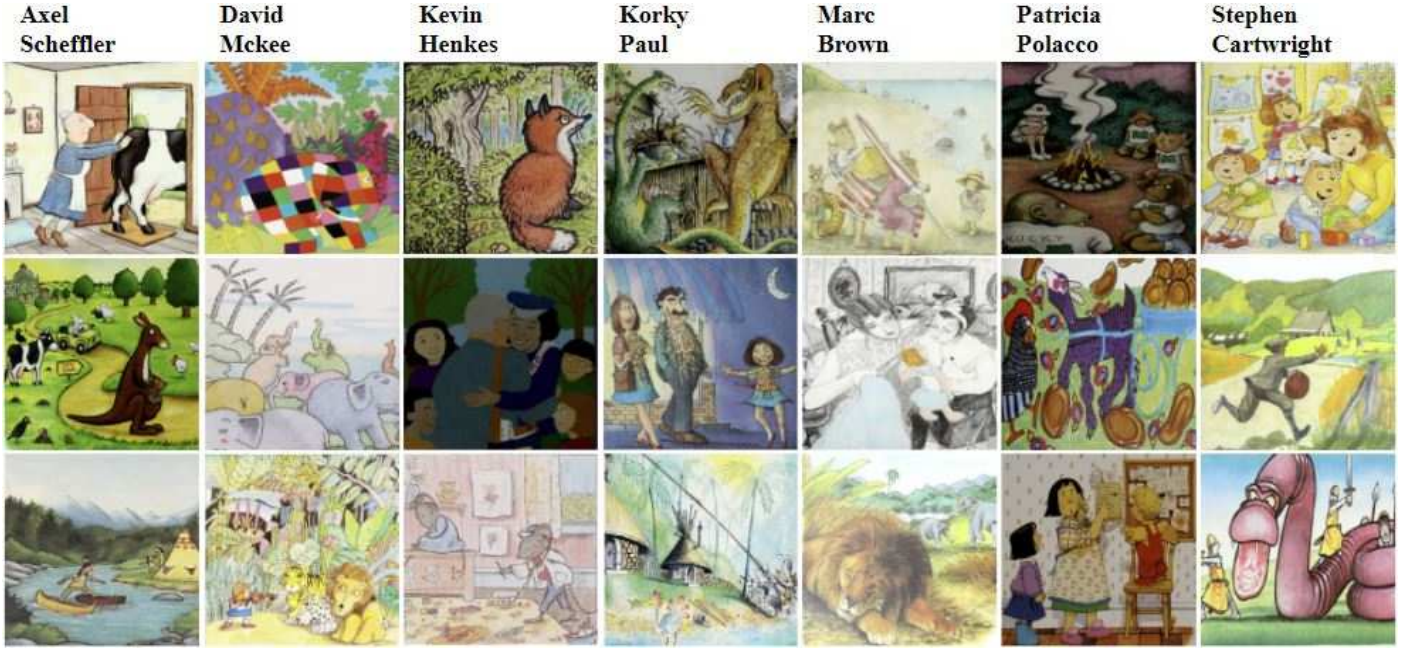}
\caption{Samples of the illustrations of the various illustrators, each drawing from a different book. Each illustrator has his own unique style of drawing. Moreover, by comparing images from the same illustrator it is notable that the drawing style may vary from one book to another (see for example Stephen Cartwright's drawings and Patricia Polacco's drawings).}
\label{fig:SampleDataset}
\end{figure*}

\paragraph{Content features loss} Inspired by Johnson et al. \cite{johnson2016perceptual}, we examined the impact of a content feature loss that measures distances in the feature space of a pretrained VGG16 network  \cite{simonyan2014very}:
\begin{equation}
\hspace{0.5in} \mathcal{L}_{content\_feat} = \mathbb{E}[||\phi(x)-\phi(y_x)||_1],
\end{equation}
where $x$ is the original content image, $y_x$ is the transformed image by the network (to the target domain), and $\phi(x)$ and $\phi(y_x)$ are the the activations of the second layer of the pretrained VGG16 network with $x$ and $y_x$ as inputs respectively. The VGG16 was pretrained on the ImageNet dataset \cite{russakovsky2015imagenet}.

Following \cite{johnson2016perceptual}, we used the 'relu2\_2' feature layer. We empirically found that the $\ell_1$ norm achieves more visually pleasing results than the $\ell_2$ norm. The addition of this feature space loss was designed to benefit the content preserving goal. 

\paragraph{Style aware content loss} Inspired by \cite{sanakoyeu2018style}, we adopted the style-aware content loss, which is an objective function that is being optimized while the network learns to transfer input images to the target domain. This objective measures the distance between the input content image $x$ and the stylized image $y_x=Decoder(Encoder(x))$ in the latent space:
\begin{eqnarray}
&& \hspace{-0.5in}\mathcal{L}_{content\_sacl} = \\ \nonumber && \hspace{-0.3in} \frac{1}{d}\mathbb{E}||Encoder(x)-Encoder(Decoder(Encoder(x)))||_2^2,
\end{eqnarray}
where the $Encoder$ and $Decoder$ are the ones used in the generator, and $d$ is the number of the latent space domains.

\paragraph{Total objective} The total objective function, which is used to train the network, is given by the following formula: 
\begin{eqnarray}
\label{eq:train_loss}
\max_D  \min_{G,F,E} \mathcal{L}_{adv}+\lambda_{sty}\mathcal{L}_{sty}-\lambda_{ds}\mathcal{L}_{ds}+\lambda_{cyc}\mathcal{L}_{cyc} \\ \nonumber
+\lambda_{feat}\mathcal{L}_{content\_feat}+\lambda_{sacl}\mathcal{L}_{content\_sacl},
\end{eqnarray}
where $\lambda_{sty}$, $\lambda_{ds}$, $\lambda_{cyc}$, $\lambda_{feat}$ and $\lambda_{sacl}$ are hyperparameters for each term, D is the discriminator, G is the generator, F represents the mapping network and E represents the style encoder of the StarGAN v2 framework (similar to \cite{choi2020stargan}). Notice that for training MISS GAN we do not use the style aware content loss (setting $\lambda_{sacl} =0$) as we have found that it does not improve performance (see ablation study in Section~\ref{sec:exp}). 

Training the proposed architecture was done using the implementation provided by the StarGan v2 authors \cite{choi2020stargan}. Namely, MISS-GAN is trained with the objective in Eq.~\ref{eq:train_loss} once using the latent vectors and once using the reference images. In the latter, the style codes are generated using the style encoder component that receives the reference images. Additional implementation details can be found in Section \ref{sec:exp}.

\section{Experiments}
\label{sec:exp}

Preserving content in image-to-image translation, yet generating styled images is a challenging task that received a wide focus recently \cite{sanakoyeu2018style, kotovenko2019content}. We now turn to show that MISS GAN achieves this successfully for image-to-illustration translation. 

\begin{figure*}[!t]
\centering
\includegraphics[width = 0.96\linewidth]{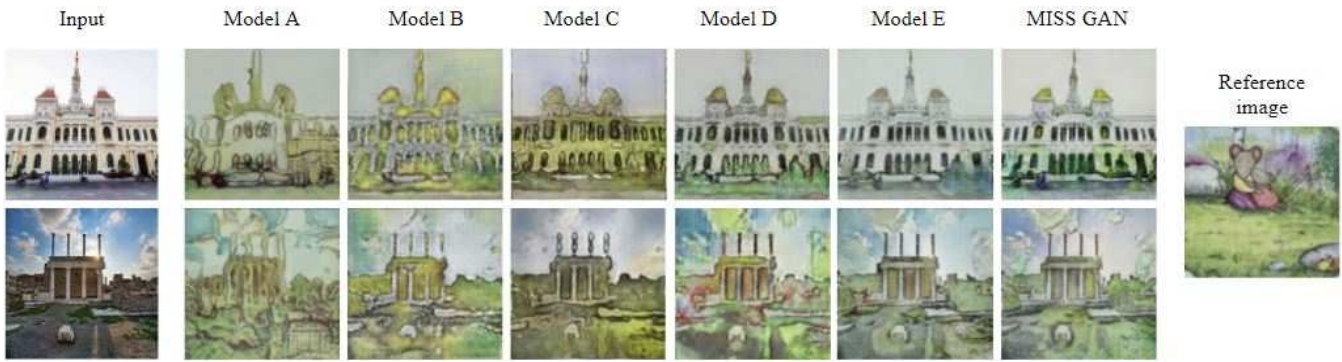}
\caption{The differences in the outputs for each ablation model (Table \ref{table:AblationsTab}). On the left, two natural input images. On the right, the reference illustration corresponding with the style code that was used to generate the images. The outputs for each ablation model are presented in the middle section.}
\label{fig:AblationRes}
\end{figure*}

\begin{figure*}[!t]
    \centering
    \includegraphics[width = 0.96\linewidth]{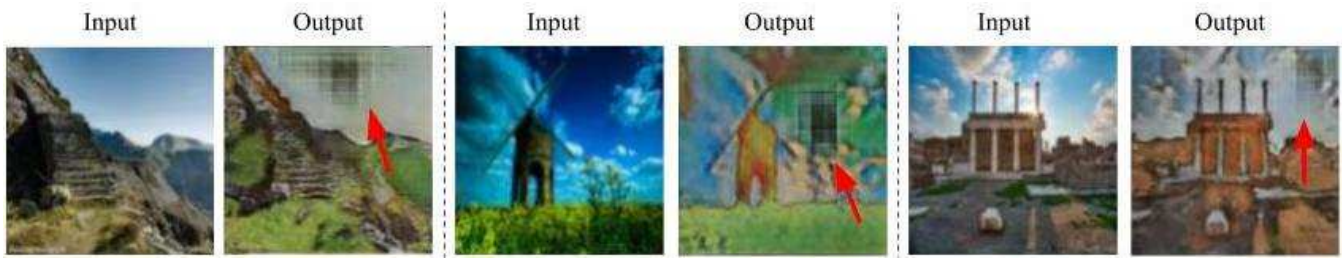}
    \caption{Additional generated images of Model B, the StarGAN v2 framework using a similar version of GANILLA generator (Fig. \ref{fig:AblationDec}). The generated outputs of Model B appeared with distortions (shown in the red arrows) that are not appealing to the human eye. Those marks did not appear while using the residual blocks in the decoder.}
  \label{fig:Disortions}
\end{figure*}

\paragraph{Dataset} For natural images, we used the dataset presented in CycleGAN \cite{zhu2017unpaired} which contains 6287 images for training and 751 images for testing. 
For the illustrations dataset, we extended 7 of the artists that were used in \cite{hicsonmez2020ganilla} and created an illustrations dataset that contains 3757 illustrations images. The number of illustrations from each illustrator is described in Table \ref{table:IlluTab}. As the dataset in \cite{hicsonmez2017draw, hicsonmez2020ganilla} was not publicly available, we re-created the dataset by scanning children's books from open source libraries (with the help of the script\footnote{https://github.com/giddyyupp/ganilla} provided by \cite{hicsonmez2020ganilla}), followed by a manual procedure of removing text areas. 

The illustrations dataset contains full-page drawings that mainly describe complex scenes. By comparing the illustrations of a specific illustrator, it is notable that the drawing style may vary from one book to another. Moreover, it can be noticed that each illustrator has his own distinct style of painting (Fig. \ref{fig:SampleDataset}). The input images to StarGAN v2 were resized to 128X128.

\begin{table}[ht]
\caption{\label{table:IlluTab}Details of the illustrations dataset}
\begin{tabular} {p{3.5cm} p{2cm} p{2cm}} 
\hline\hline
Illustrator & Book Count & Image Count \\ [0.5ex] 
\hline
Axel Scheffler & 15 & 426 \\ 
David Mckee & 21 & 548  \\
Kevin Henkes & 18 & 400  \\
Korky Paul & 18 & 470  \\
Marc Brown & 20 & 522  \\ 
Patricia Polacco & 28 & 866  \\
Stephen Cartwright & 22 & 525  \\
\hline
\end{tabular}
\end{table}

\begin{figure}[!t]
\centering
\includegraphics[scale=.58]{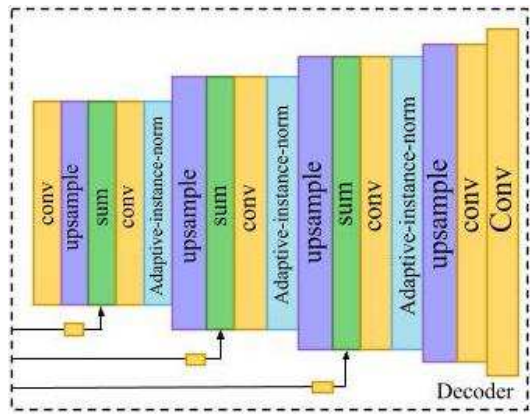}
\caption{Overview of the decoder of the ablation model, an architecture that resembles more the GANILLA generator architecture \cite{hicsonmez2020ganilla}, with additional convolution and AdaIN layers in order to integrate AdaIN information. Note that the three arrows with the small yellow rectangle represent the skip connections originating from the encoder (as in Fig. \ref{fig:MISSGANGen}).}
\label{fig:AblationDec}
\end{figure}

\paragraph{Implementation details} 
Our models were trained on all of the images in the illustrations dataset. The number of domains was defined to be two (one domain for illustrations and another for natural images). Moreover, all models were trained from scratch. Note that we trained our models with 2,000,000 iterations and all images were resized to 128$\times$128 pixels. We used GeForce GTX 1070 with 8GB with which it took approximately two days to train each model. Training with larger images, 256$\times$256 as suggested by Choi et al.\cite{choi2020stargan}, may further improve results. Our code is available online \footnote{https://github.com/NoaBrazilay/MISSGAN}.

\begin{figure*}[!t] 
    \centering
    \includegraphics[width = 1\linewidth]{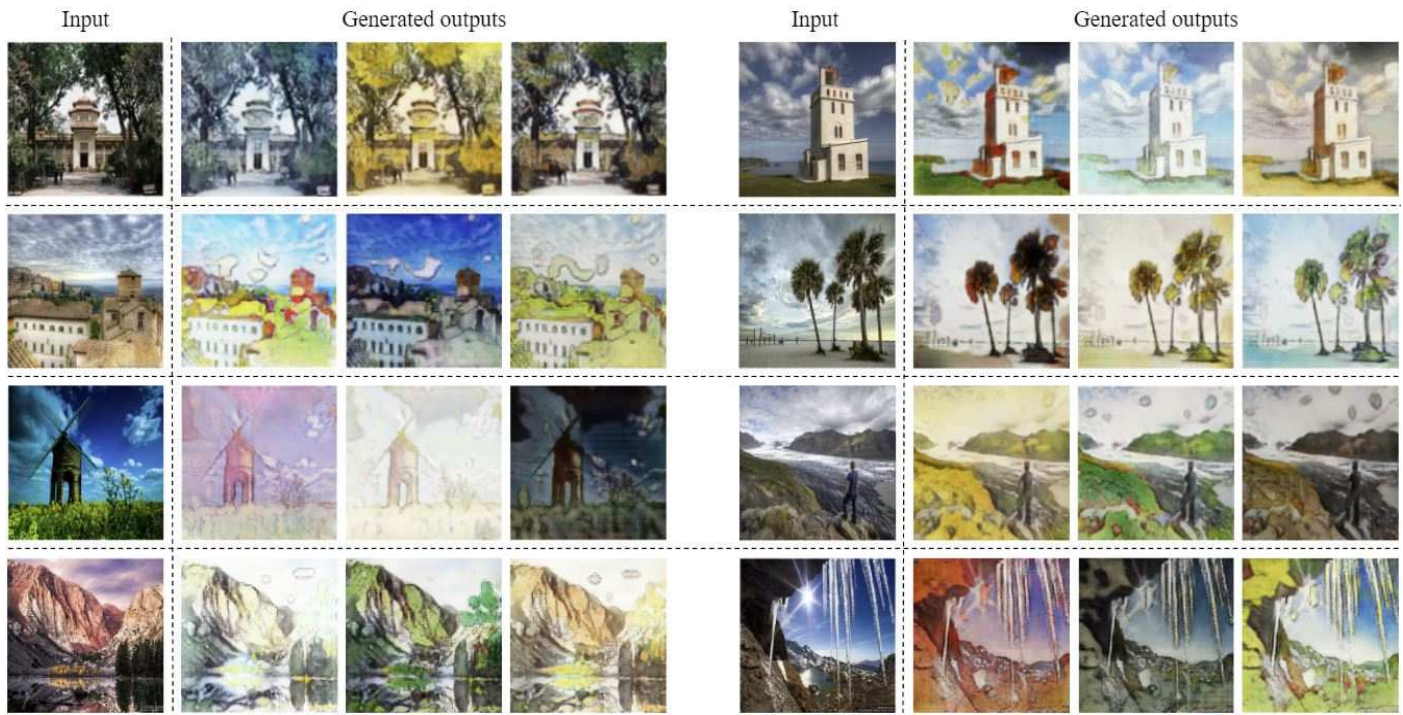}
    \caption{Different generated illustrations of MISS GAN for several natural input images. The output illustrations were generated using style codes that were created by the Mapping network component.}
  \label{fig:MappingNetRes}
\end{figure*}

\begin{figure*}[!t] 
    \centering
    \includegraphics[width = 1\linewidth]{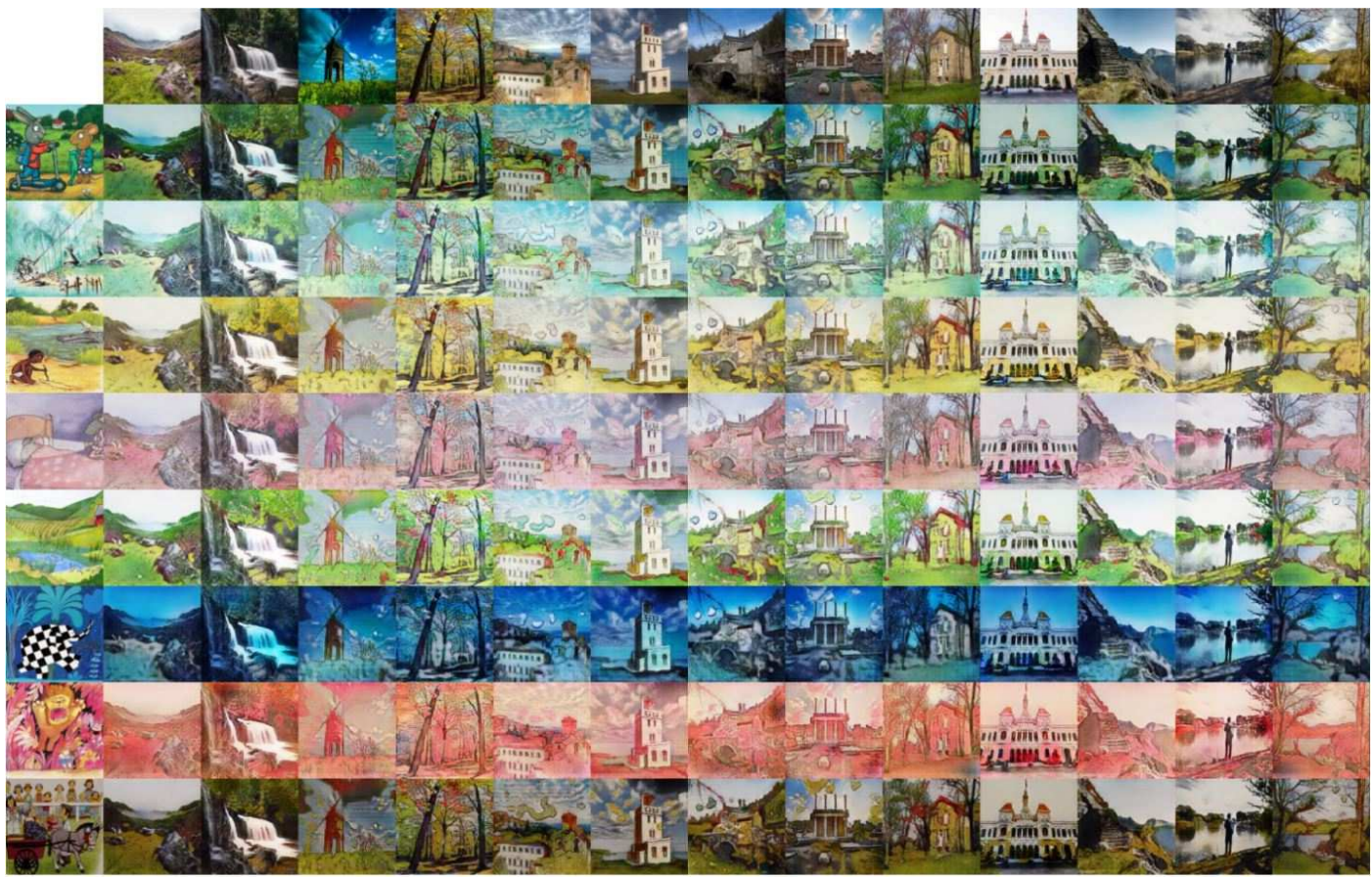}
    \caption{Different generated illustrations of MISS GAN for several natural input images while using reference illustrations. Those images were generated using the style code created by the style encoder component according to the reference illustrations.}
  \label{fig:ResStyleEnc}
\end{figure*}

\paragraph{Comparison to StarGAN and ablation}
To check the effect of each of the components of our model we perform an ablation study in which we compare MISS GAN to the baseline StarGAN v2 and check the impact of the added residual connections in the decoder and different loss functions.
We empirically examined the performances of four other ablation models (Table \ref{table:AblationsTab}). 
The first ablation Model, Model B, examines the effect of replacing the StarGAN v2 generator with an architecture that resembles the one in \cite{hicsonmez2020ganilla}. The generator is composed of an encoder similar to  \cite{hicsonmez2020ganilla}, and a decoder to which we have added AdaIN layers and convolutional layers to integrate style information (Fig. \ref{fig:AblationDec}). 

The second Model, Model C, is similar to Model B, however, it contains residual blocks in the generator's decoder (as in Fig. \ref{fig:MISSGANGen}). Model D is similar to model C, however, it was trained with an additional objective function, the style-aware content loss \cite{sanakoyeu2018style}. Model E is basically equivalent to Model D, except of an additional objective content loss function, the features content loss. The proposed MISS GAN is similar to Model C, however, it uses the feature content loss as an additional loss function.

We used the StarGAN v2 framework with no modifications as our baseline model. We then trained it on the two datasets while using the implementation provided by the authors, however, the learning rates of the discriminator, generator, and style encoder components were set to $10^{-4}$.


\begin{table}[!t]
\caption{\label{table:AblationsTab}Details of the trained various models}
\centering
\begin{tabular} {p{1.65cm} p{6.2cm}} 
\hline\hline
Model & Method \\ [0.5ex] 
\hline
A & Baseline StarGAN V2 \cite{choi2020stargan}  \\ 
B & + A version of GANILLA generator
(Fig. \ref{fig:AblationDec}) \\
C & + Res-blocks in generator's decoder (Fig.  \ref{fig:MISSGANGen})\\
D & + Style aware content loss  \\
E & + Features content loss \\ 
MISS GAN & Model C + features content loss\\
\hline
\end{tabular}
\end{table}

Fig. \ref{fig:AblationRes} shows the visual results of the different models. By observing the different generated outputs, it is apparent that using the GANILLA generator significantly improves the results by preserving better content information (Model B compared to the baseline). The generated outputs of the baseline model, Model A, are distorted, and content preservation hardly occurs. 

Another modification that benefits content preservation is adding the residual blocks in the generator's decoder (Fig. \ref{fig:MISSGANGen}). By comparing the generated outputs of Model B and Model C, it appears that Model C generates styled images and yet preserves more edges derived from the input image (for example, compare the windows of the buildings between the two generated images in Fig. \ref{fig:AblationRes}). Adding the residual blocks in the decoder also eliminates unappealing distortions that appears in the generated images of Model B (Fig. \ref{fig:Disortions}). 

Significantly, the baseline model and the two first ablation models do not preserve the content of the input image properly. Therefore, we adopted two content loss functions that were proposed in \cite{johnson2016perceptual, sanakoyeu2018style}. We inspected the performances of the framework with these additional content loss functions (Model D, Model E and MISS GAN). Fig. \ref{fig:AblationRes} presents the results. By observing the generated illustrations, it appears that using the feature content objective with the proposed architecture benefits the most. The proposed MISS GAN is able to create appealing yet content preserving illustrations with diverse styles.

\paragraph{Style code based generation} More generated illustrations are presented in Fig. \ref{fig:MappingNetRes}. Those images were generated using style codes that were created by the mapping network component. It is very noticeable that the generated outputs are different in style. Particularly, by observing Fig. \ref{fig:MappingNetRes}, it is evident that various diverse styles and color schemes are present among the generated outputs (darker and brighter schemes which appear in the illustrations dataset, as shown in Fig. \ref{fig:SampleDataset}).

\paragraph{Reference illustration based generation} Illustrations that were generated using a style code that was created by the style encoder component are presented in Fig. \ref{fig:ResStyleEnc}. By observing the generated illustrations in this figure, it is possible to see the great effect each reference illustration has on its related generated images. Since our model was trained on all of the illustrators' images, it utilizes this information to generate a styled illustration that is fine-tuned according to a reference image which correlates to a specific illustrator.

\begin{figure}[!t] 
    \centering
    \includegraphics[scale= 0.54]{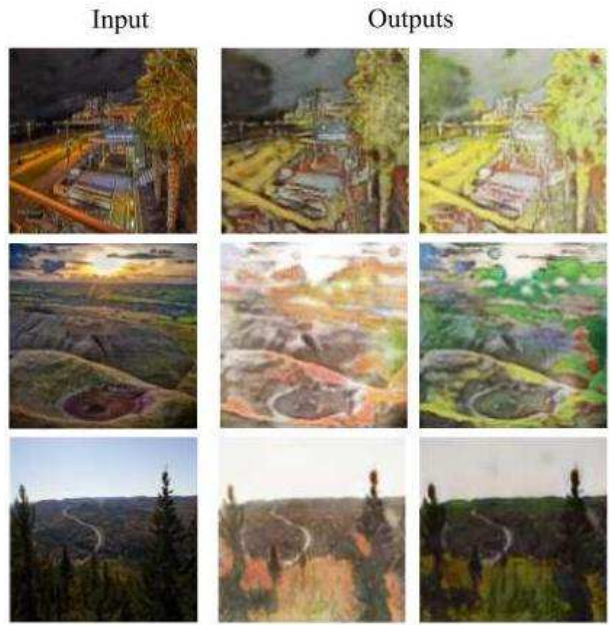}
    \caption{Examples of failure cases of MISS GAN}
  \label{fig:MISSFail}
\end{figure}

\paragraph{MISS GAN failures} The illustrations dataset which we used to train MISS GAN contains highly abstract drawings. Therefore, transformations of natural images that contain small shapes and details are lacking some of the high-frequency details that are present in the original image. Samples of failure cases are shown in Fig. \ref{fig:MISSFail}.

\section{Conclusions}
In this work, we proposed the MISS GAN framework, which successfully transfers natural images to diverse styled illustrations using only one trained generator. The generated images preserved the content of the input image while altering its style according to a reference illustration or according to a randomized latent code. Existing architectures require to train several generators to handle the variety of styles that can be found in the illustrations dataset \cite{hicsonmez2020ganilla}. Moreover, performing the ablation study emphasized the importance of the architectural modifications we performed compared to the baseline model.

The evaluation in this work was performed by using empirical evaluation. Some works utilize the FID \cite{heusel2017gans} and the Inception score \cite{salimans2016improved} for evaluating GANs performance. Yet, such measures were not proved as efficient in the illustration field. For future work, subjective human scoring or other automatic methods may be incorporated in the evaluation stage.

There are several possible future directions to this work.
First, all images have been reduced to 128$\times$128 pixels. Training on larger images may further improve results. This may be achieved by training our generator in a progressive way as done in \cite{karras2018progressive}.
Also, since the illustrations dataset contains highly abstract drawings, it appears that natural images that contain rich information of high frequencies fail to transfer properly. This might be mitigated by adding a loss on the higher frequencies. 
Recent papers have shown the power of working with the Fourier domain with deep neural networks \cite{tesfaldet2019fourier}. Thus, to address the lack of high frequency details in the generated images, it might be possible to compute the feature space objectives in the Fourier domain, which is likely to mitigate this problem. Finally, while this work focused on illustrations, we believe that the proposed MISS GAN can be used to handle also other types of multi-illustrator/designer style images.

\bibliographystyle{elsarticle-num}
\bibliography{refs}

\end{document}